\documentclass[aps,prl,superscriptaddress,twocolumn,nopacs,amsmath,amssymb,letter]{revtex4-1}

\usepackage{color}
\usepackage{graphicx}
\usepackage{dcolumn} 
\usepackage{bm}
\usepackage{xspace}
\setcitestyle{numbers,square}

\begin{document}

\title{Fermiology and Superconductivity of Topological Surface States in PdTe$_2$}

\author{O.~J.~Clark}
\author{M.~J.~Neat}
\affiliation {SUPA, School of Physics and Astronomy, University of St. Andrews, St. Andrews KY16 9SS, United Kingdom}

\author{K.~Okawa}
\affiliation{Materials and Structures Laboratory, Tokyo Institute of Technology, Kanagawa 226-8503, Japan}

\author{L.~Bawden}
\affiliation {SUPA, School of Physics and Astronomy, University of St. Andrews, St. Andrews KY16 9SS, United Kingdom}

\author{I.~Markovi{\'c}}
\affiliation {SUPA, School of Physics and Astronomy, University of St. Andrews, St. Andrews KY16 9SS, United Kingdom}
\affiliation {Max Planck Institute for Chemical Physics of Solids, N{\"o}thnitzer Stra{\ss}e 40, 01187 Dresden, Germany}

\author{F.~Mazzola}
\affiliation {SUPA, School of Physics and Astronomy, University of St. Andrews, St. Andrews KY16 9SS, United Kingdom}

\author{J.~Feng}
\affiliation {SUPA, School of Physics and Astronomy, University of St. Andrews, St. Andrews KY16 9SS, United Kingdom}
\affiliation {Suzhou Institute of Nano-Tech. and Nanobionics (SINANO), CAS, 398 Ruoshui Road, SEID, SIP, Suzhou, 215123, China}

\author{V.~Sunko}
\affiliation {SUPA, School of Physics and Astronomy, University of St. Andrews, St. Andrews KY16 9SS, United Kingdom}
\affiliation {Max Planck Institute for Chemical Physics of Solids, N{\"o}thnitzer Stra{\ss}e 40, 01187 Dresden, Germany}

\author{J.~M.~Riley}
\affiliation {SUPA, School of Physics and Astronomy, University of St. Andrews, St. Andrews KY16 9SS, United Kingdom}
\affiliation{Diamond Light Source, Harwell Campus, Didcot, OX11 0DE, United Kingdom}

\author{W.~Meevasana}
\affiliation{School of Physics and Center of Excellence on Advanced Functional Materials, Suranaree University of Technology, Nakhon Ratchasima, 30000, Thailand}
\affiliation{ThEP, Commission of Higher Education, Bangkok 10400, Thailand}

\author{J.~Fujii}
\author{I.~Vobornik}
\affiliation{Istituto Officina dei Materiali (IOM)-CNR, Laboratorio TASC, in Area Science Park, S.S.14, Km 163.5, I-34149 Trieste, Italy}

\author{T.~K.~Kim}
\author{M.~Hoesch}
\affiliation{Diamond Light Source, Harwell Campus, Didcot, OX11 0DE, United Kingdom}

\author{T.~Sasagawa}
\affiliation{Materials and Structures Laboratory, Tokyo Institute of Technology, Kanagawa 226-8503, Japan}

\author{P. Wahl}
\email{wahl@st-andrews.ac.uk}
\affiliation {SUPA, School of Physics and Astronomy, University of St. Andrews, St. Andrews KY16 9SS, United Kingdom}

\author{M.~S.~Bahramy}
\email{bahramy@ap.t.u-tokyo.ac.jp}
\affiliation{Quantum-Phase Electronics Center and Department of Applied Physics, The University of Tokyo, Tokyo 113-8656, Japan}
\affiliation{RIKEN center for Emergent Matter Science (CEMS), Wako 351-0198, Japan}

\author{P.~D.~C.~King}
\email{philip.king@st-andrews.ac.uk}
\affiliation {SUPA, School of Physics and Astronomy, University of St. Andrews, St. Andrews KY16 9SS, United Kingdom}

\date{\today}

\begin{abstract}
{We study the low-energy surface electronic structure of the transition-metal dichalcogenide superconductor PdTe$_2$ by spin- and angle-resolved photoemission, scanning tunneling microscopy, and density-functional theory-based supercell calculations. Comparing PdTe$_2$ with its sister compound PtSe$_2$, we demonstrate how enhanced inter-layer hopping in the Te-based material drives a band inversion within the anti-bonding $p$-orbital manifold well above the Fermi level. We show how this mediates spin-polarised topological surface states which form rich multi-valley Fermi surfaces with complex spin textures. Scanning tunneling spectroscopy reveals type-II superconductivity at the surface, and moreover shows no evidence for an unconventional component of its superconducting order parameter, despite the presence of topological surface states.}
\end{abstract}

\maketitle

There has long been interest in systems where spin-polarised electronic states co-exist with superconductivity~\cite{read_paired,Gorkov,schnyder_classifcation, sato_topological, sun_dirac,sato_topological2}. This is thought as a promising route to generating a spin-triplet component of the superconducting order parameter,  and to realise topological superconductors which host Majorana zero modes at their boundaries or in vortex cores~\cite{sato_topological, sato_topological2}. Such conditions are proposed to be realised if superconductivity can be induced in the spin-momentum locked surface states of materials hosting non-trivial bulk band topology. There have been extensive efforts to achieve proximity-coupled superconductivity in topological surface states by interfacing with conventional superconductors \cite{maier_induced, deacon_jopheson, Zareapour_proximity_2012,yang_proximity_2012, wang_fully_2013} as well as to induce superconductivity by extrinsic doping of bulk topological insulators~\cite{hor_superconductivity,matano_spin,levy_experimental}. The latter approach, however, often suffers from the difficulty of achieving high superconducting volume fractions~\cite{hor_superconductivity,Schneeloch_CuBi2Se3}, motivating the search for materials which host topologically non-trivial states and are simultaneously robust superconductors. 

In this Letter, we demonstrate that pristine PdTe$_2$, an intrinsic bulk superconductor ($T_c\approx1.7$~K) \cite{raub_occurrence,kjekshus_constitution,ryu_superconductivity}, hosts topological surface states which intersect the Fermi level. We demonstrate, from spin- and angle-resolved photoemission spectroscopy (spin-ARPES) and density-functional theory (DFT), that these form complex multi-pocket Fermi surfaces with intricate spin textures. Nonetheless, tunneling measurements into the surface layer from scanning tunneling microscopy and spectroscopy (STM/STS) indicate the superconductivity at the surface is of a conventional $s$-wave form consistent with Bardeen-Cooper-Schrieffer (BCS) theory.

\begin{figure*}
\includegraphics[width=\textwidth]{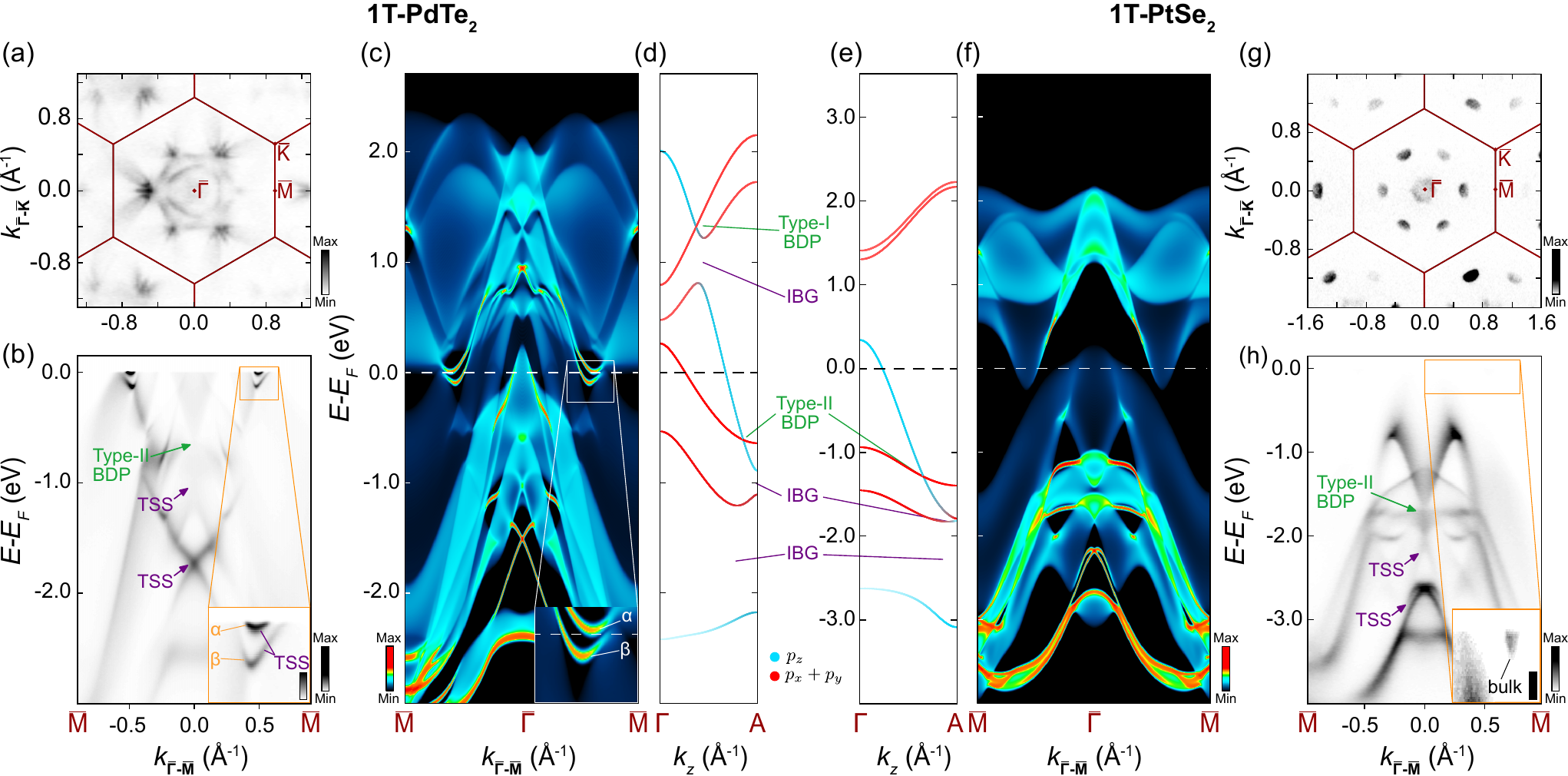}
\caption{ \label{f:overview} (a) ARPES Fermi surface of PdTe$_2$ ($E_F\pm{5}$~meV; $h\nu=107$~eV; p-pol, probing close to an A-plane along $k_z$). (b) In-plane dispersion along the $\overline{\Gamma}-\overline{\mbox{M}}$ direction ($h\nu=24$~eV, probing a similar $k_z$ value as in (a)). (c) Corresponding supercell calculations of the dispersion over an extended energy range, projected onto the top two unit cells of PdTe$_2$. (d) Out-of-plane bulk band dispersions along the $\Gamma-\mbox{A}$ direction of the Brillouin zone from DFT calculations, projected onto the chalcogen $p_{x,y}$ (red) and $p_z$ (cyan) orbitals. (e-h) Equivalent measurements and calculations for PtSe$_2$ (ARPES measurements: $h\nu=107$~eV; p-pol). Topological surface states (TSS), bulk Dirac points (BDP), and inverted band gaps (IBG) are labeled.}
\end{figure*}

Single-crystal PdTe$_2$ and PtSe$_2$ samples (space group: $P\bar{3}m1$) were cleaved {\it in situ} at temperatures of $\sim\!10$~K. Spin-resolved ARPES measurements were performed at the APE beamline of Elettra Sincrotrone Trieste using a Scienta DA30 analyser fitted with very low energy electron diffraction (VLEED) based spin polarimeters \cite{vleed}.  Spin-integrated measurements were performed at the I05 beamline of Diamond Light Source using a Scienta R4000 hemispherical analyser \cite{R4000}. $p$-polarised light and a measurement temperature of 5-15~K was used throughout. STM measurements were performed using a home-built system operating in cryogenic vacuum \cite{singh_construction,white_stiff}. Measurements were performed in a magnetic field up to 1T, and the sample temperature was held constant at $\sim\!40$~mK (superconducting state) or $8$~K (normal state). STM tips were cut from a platinum/iridium wire. Bias voltages were applied to the sample with the tip at virtual ground. Differential conductance spectra were recorded through a standard lock-in technique with frequency $f=437$~Hz. A superconducting tip was obtained by collecting a small piece of the sample at its apex. Fully-relativistic DFT calculations were performed using the Perdew-Burke-Ernzerhof exchange-correlation functional as implemented in the WIEN2K program \cite{wien2k}, using a $20\times 20\times 20$ $k$-mesh. To calculate the surface electronic structure, tight binding supercells containing 100 formula units of PdTe$_2$ or PtSe$_2$ stacked along the crystalline $c$-direction were constructed from the bulk DFT calculations using maximally localized Wannier functions with chalcogen $p$-orbitals and transition-metal $d$-orbitals as the projection centres \cite{souza,mostofi,kunes}.  

The electronic structure of PdTe$_2$, as well as its sister compound PtSe$_2$, are summarised in Fig.~\ref{f:overview}. While the Pd/Pt $d$ states retain a fully-filled configuration, and are thus located well below the Fermi level~\cite{chhowalla_2013}, a conducting state is obtained due to an energetic overlap of predominantly Te/Se-derived bonding and anti-bonding states~\cite{furuseth_redetermined}. In PdTe$_2$, this leads to a complex multi-band Fermi surface (Fig.~\ref{f:overview}(a)). Strong inter-layer interactions for the chalcogen-derived states render this Fermi surface three-dimensional, despite the layered nature of this compound. Many of the observed spectral features are therefore diffuse in our measured Fermi surface from ARPES, reflecting the inherent surface sensitivity, and thus poor $k_z$ resolution, of photoemission. Similarly broad features, including a recently-identified type-II bulk Dirac cone ~\cite{bahramy_ubiquitous, noh_experimental_2017} centered at $E-E_F=-0.65$~eV, can also be seen in our measured dispersions (Fig.~\ref{f:overview}(b)) as well as in our supercell calculations where bulk bands at different $k_z$ values are projected onto the surface plane (Fig.~\ref{f:overview}(c). 

A number of much sharper features are also observed. A Dirac cone situated $\sim\!1.75$~eV below the Fermi level is clearly evident in our ARPES. This has recently been identified as a topological surface state in PdTe$_2$ and related compounds \cite{yan_identification_2015,yan_lorentz-violating_2016,bahramy_ubiquitous,huang_type-ii_2016,zhang_experimental_2017}. This arises from a band inversion that occurs within the Te $p$-orbital manifold, and is induced by a naturally disparate out-of-plane dispersion of $p_z$ and $p_{x/y}$-derived bands along $k_z$ (Fig.~\ref{f:overview}(d))~\cite{bahramy_ubiquitous}. This same mechanism drives the formation of both the type-II bulk Dirac cone and a further topological surface state located $\sim\!1$~eV below $E_F$, that is visible in our supercell calculations presented here (Fig.~\ref{f:overview}(c)) and is clearly resolved experimentally in Ref.~\cite{bahramy_ubiquitous}. These topological states, however, are too far below the Fermi level to play any role in the superconductivity of this system. On the other hand, our measurements (Fig.~\ref{f:overview}(b), see inset) and calculations (Fig.~\ref{f:overview}(c)) reveal an additional pair of sharp spectral features which intersect the Fermi level approximately mid-way along the $\overline{\Gamma}-\overline{\mbox{M}}$ direction. These have negligible dispersion in the out-of-plane direction (Supplementary Fig.~S1~\cite{sup}), and we thus assign them as surface states. 

Their origin is evident from the calculated bulk band structure shown along the out-of-plane direction in Fig.~\ref{f:overview}(d). In PdTe$_2$, the inter-layer hopping between $p_z$ orbitals in neighbouring layers is sufficiently high that the $p_z$-derived band crosses through both the bonding and anti-bonding $p_{x/y}$-derived bands (lower and upper pair, respectively, of the red coloured bands in Fig.~\ref{f:overview}(d)) as they disperse along the $k_z$ direction. A protected bulk Dirac point and an inverted band gap with non-trivial $\mathbb{Z}_2$ topological order are therefore generated above the Fermi level. The inverted band gap above $E_F$ should give rise to a topological surface state. While this band inversion is $\sim\!1$~eV above the Fermi level along $\Gamma-\mbox{A}$, the in-plane bandwidths are large, and the relevant band inversion can be traced to the bulk states in the vicinity of the Fermi level at the time-reversal invariant $\overline{\mbox{M}}$-point. Here, the pair exchange arising from the bulk boundary correspondence of topological surface states~\cite{Bahramy_pair,Wray} enforces that the dispersion of the topological surface states are pulled down towards the Fermi level. Experimentally, we find that both the ``upper'' (labeled $\alpha$) and ``lower'' (labeled $\beta$) branch of the topological state cross $E_F$ (Fig.~\ref{f:overview}(b, inset)), although the occupied bandwidth of the $\alpha$-band along this direction is small ($\approx20$~meV). 

To validate the above picture, we compare PdTe$_2$ with its sister compound PtSe$_2$. The occupied electronic structure is similar to PdTe$_2$, hosting a type-II bulk Dirac cone and a pair of topological surface states below $E_F$ (Fig.~\ref{f:overview}(f,h)). These arise from the crossing of the Se $p_z$-derived band with the bonding Se $p_{x/y}$ bands (Fig.~\ref{f:overview}(e))~\cite{bahramy_ubiquitous,huang_type-ii_2016}. The bandwidth of the Se $p_z$-derived band is, however, much smaller than that of the Te-derived one owing to the higher electronegativity of Se than Te. This leads to a stronger metal-chalcogen bond, and therefore weaker interlayer hopping in PtSe$_2$. Crucially, although a semi-metallic ground state still occurs (Fig.~\ref{f:overview}(g)), the upper and lower branches of Se-derived states remain completely separable in PtSe$_2$. The band inversions above $E_F$, and resulting topological surface state crossing the Fermi level, that occur for PdTe$_2$ are therefore absent in this compound (Fig.~\ref{f:overview}(e-h)). 

\begin{figure}
\includegraphics[width=\columnwidth]{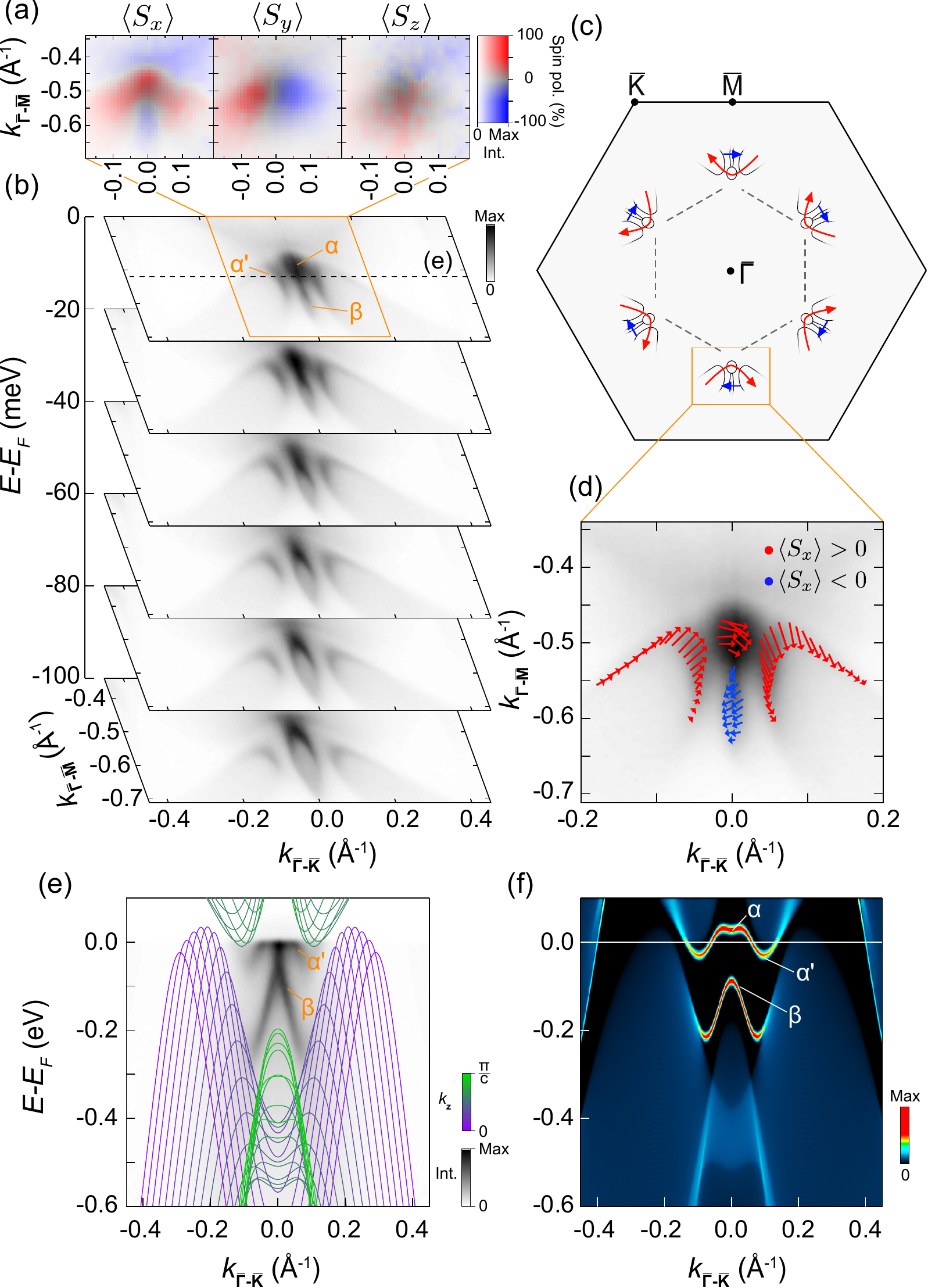}
\caption{ \label{f:spin} (a) Three-component spin-resolved ARPES Fermi surface measured ($h\nu=24eV$) over the range shown in (b). Here  $\langle S_x \rangle$ and $\langle S_y \rangle$ are perpendicular to and along the $\overline{\Gamma}-\overline{\mbox{M}}$ direction, respectively. $\langle S_z \rangle$ is the out-of-plane spin component. (b) Near-$E_F$ constant energy contours measured from spin-integrated ARPES ($h\nu=24eV$) over the portion of the surface Brillouin zone where the topological states reside. (c) Schematic representation of the global Fermi surface spin texture throughout the surface Brillouin zone that is deduced from our measurements. (d) The in-plane spin texture (arrows) determined directly from the spin-resolved measurements and shown atop the measured spin-integrated Fermi surface segment. (e) ARPES-measured dispersion ($h\nu=24$~eV) and (f) corresponding supercell calculation, cutting through the TSS Fermi surfaces along the dashed line indicated in (b). $k_z$-dependent bulk band calculations are overlaid in (e) and visible as diffuse spectral weight in (f), demonstrating how small projected band gaps control the dispersion of the TSS (see also Supplemental Fig.~S2~\cite{sup}). 
}
\end{figure}

Figure~\ref{f:spin} shows the Fermi surfaces formed by the topological states which intersect the Fermi level in the Te-based material. The $\alpha$ band forms a small nearly circular electron pocket, located approximately midway along $\overline{\Gamma}-\overline{\mbox{M}}$. This electron pocket rapidly shrinks and then vanishes in constant energy slices taken below the Fermi level (Fig.~\ref{f:spin}(b)), reflecting the narrow occupied bandwidth of this band along $\overline{\Gamma}-\overline{\mbox{M}}$. However, this is only a local minimum of the surface state dispersion located within a narrow projected bulk band gap along this direction (see Fig.~\ref{f:spin}(e,f); further cuts are shown in Supplemental Fig.~S2~\cite{sup}). The band disperses upwards above the Fermi level along the direction perpendicular to $\overline{\Gamma}-\overline{\mbox{M}}$ before turning over again to form the intense rim of spectral weight that borders the three-dimensional bulk bands that are evident as diffuse filled-in spectral weight spanning out away from the $\overline{\Gamma}-\overline{\mbox{M}}$ line in Fig.~\ref{f:spin}(b). With increasing momentum away from this line, these surface states become degenerate with the bulk bands, and ultimately lose all spectral weight. They thus appear to form open-ended arc-like features. We stress that these are not Fermi arcs of the form discussed extensively as spanning surface projections of bulk Dirac and Weyl points
\cite{young_dirac_2012,borisenko_experimental_2014,liu_discovery_2014,wan_topological_2011,xu_discovery_2015,lv_observation_2015,huang_spectroscopic_2016,soluyanov_type-ii_2015,mccormick_minimal_2017}. Instead, they reflect the small projected bulk band gaps here (Fig.~\ref{f:spin}(b) and Supplemental Fig.~S2~\cite{sup}), with the surface states merging into the bulk continuum as they disperse away from the high-symmetry line. As these surface features derive from the same state as the $\alpha$ pocket, we label them $\alpha'$ in Fig.~\ref{f:spin}.

The small projected band gaps of the bulk spectrum thus lead to only small momentum space regions centered along the $\overline{\Gamma}-\overline{\mbox{M}}$ directions in which the surface states are well defined. This results in a rich multi-valley Fermi surface (Fig.~\ref{f:overview}(a)), far removed from the generic isolated near-circular Fermi surfaces of topological surface states in, e.g., Bi$_2$Se$_3$ class topological insulators. Nonetheless, we show in Fig.~\ref{f:spin} that the Fermi surfaces observed here still maintain a spin texture reminiscent of the simple chirality of a conventional topological surface state. 

Our spin-ARPES measurements reveal that the surface states host a strong spin polarisation ($>$ 70\% from fits to energy distribution curves (EDCs); Supplemental Fig.~S3~~\cite{sup}). Along $\overline{\Gamma}-\overline{\mbox{M}}$, the chiral $\langle S_x \rangle$ (perpendicular to $\overline{\Gamma}-\overline{\mbox{M}}$) spin component is dominant (Fig.~\ref{f:spin}(a)), consistent with the underlying trigonal symmetry of the crystal surface. The $\alpha$ and $\beta$ bands have opposite spin polarisation, supporting their assignment as the two branches of a topological surface state. Both Fermi crossings of the $\alpha$ band along $\overline{\Gamma}-\overline{\mbox{M}}$ have the same sign of $\langle S_x \rangle$. The spin, therefore, does not wind around the closed circular $\alpha$-band Fermi surface. Instead, it develops a finite $\langle S_y \rangle$ (parallel to $\overline{\Gamma}-\overline{\mbox{M}}$) component of opposite sign on the two sides of the high-symmetry $\overline{\Gamma}-\overline{\mbox{M}}$ line, canting the spin towards the Brillouin zone boundary in the in-plane direction. Similarly, the two $\alpha'$-bands have the same sign of $\langle S_x \rangle$ but opposite sign of $\langle S_y \rangle$, as shown extracted from our experimental spin-polarised Fermi surface maps in Fig.~\ref{f:spin}(d). The overall result is a Fermi surface spin-texture of the $\alpha$-derived Fermi surfaces that has a global chiral winding around the Brillouin zone centre, but with a significant radial component developing away from the $\overline{\Gamma}-\overline{\mbox{M}}$ line (Fig.~\ref{f:spin}(c)), similar to what might be expected for a more conventional topological surface state if it develops a hexagonal warping away from circular geometry \cite{fu_warping,michiardi_strongly_2015}. While at the Fermi level only the chiral component of the $\beta$-band is visible, similar radial components, as well as an out-of-plane spin canting, develop for its highly-fragmented surface state contours below $E_F$ (Supplemental Fig.~S4~~\cite{sup}).

\begin{figure}
\includegraphics[width=\columnwidth]{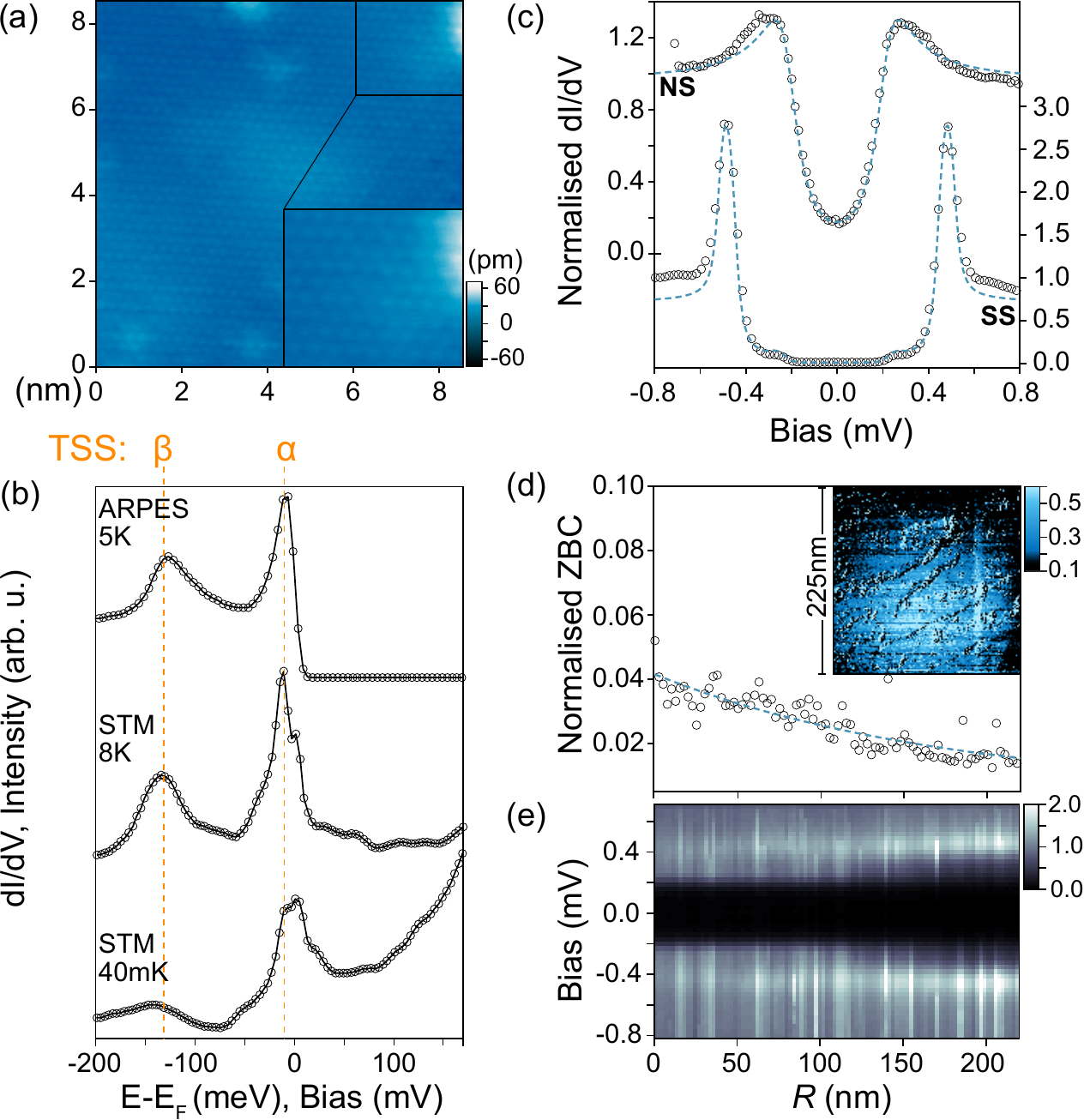}
\caption{ \label{f:STM} (a) Surface topography of PdTe$_2$ measured at $T=8$~K from a $8.5\times 8.5$~nm$^2$ region (Bias voltage: $V=10$~mV, Tunneling current set-point: $I=0.4$~nA). (b) ARPES spectra (top; $T=5$~K, $h\nu=24$~eV, integrated along the $\overline{\Gamma}-\overline{\mbox{M}}$ direction of the Brillouin zone) and differential conductance spectra measured using STM at $T=8$~K (middle) and $T=40$~mK (bottom). (c) Low-energy spectra measured at $40$~mK and performed with a normal (N, top) and superconducting (S, bottom) tip. The NS-junction (SS-junction) spectra are averaged over a 5.4 x 5.4~nm$^2$ (10$\times$10~nm$^2$) area and are normalised to the conductance at 0.6~mV. 
The dashed lines show the result of Dynes fits, incorporating thermal broadening of 100~mK \cite{singh_construction}. For the NS-junction, a lock-in amplitude of 25~$\mu$V was used, with the fit yielding a sample superconducting gap size of $\Delta=215~\mu$V with a Dynes broadening parameter of $\Gamma=65~\mu$V. For the SS-junction, a lock-in amplitude of $30~\mu$V was used with Dynes broadenings of 18 $~\mu$V and $4~\mu$V for sample and tip respectively. The obtained superconducting gap of the sample and tip is 240$~\mu$V. (d) Radially averaged decay of zero bias conductance (ZBC) with distance ($R$) from the centre of a vortex core, measured with a superconducting tip in a magnetic field of 7$\pm$2~mT. The inset shows the real-space image of the vortex via its enhanced ZBC. (e) The radial dependence of the full superconducting gap structure with distance from the centre of the vortex core, obtained using V=8~mV, I=0.4~nA and a lock in amplitude of 30~$\mu$V. }
\end{figure}

These findings raise an exciting prospect to investigate how such topological states, with their complex Fermi surface spin textures, interplay with the bulk superconductivity of PdTe$_2$. To this end, we investigate the superconductivity at the surface using low-temperature STM (Fig.~\ref{f:STM}). Our measured tunneling spectra at 8~K (Fig.~\ref{f:STM}(b)) reveal two pronounced peaks in the local density of states centered at approximately -10~meV and -130~meV. These are in excellent agreement with the positions of clear peaks in the density of states arising from the van Hove singularities of the $\alpha$ and $\beta$ TSS bands, evident in our angle-integrated ARPES spectra (Fig.~\ref{f:STM}(b)). Such features remain in our STS measurements upon cooling (Fig.~\ref{f:STM}(b)), indicating the persistence of the Fermi-level TSSs observed here into the superconducting state. Despite this, our low-temperature tunneling spectra (Fig.~\ref{f:STM}(c)) reveal a clearly-resolved U-shaped superconducting gap, indicating nodeless superconductivity. This is well described by a Dynes model fit, indicating a fully-gapped state with a superconducting gap size $\Delta=215\pm{2}~\mu$V. This gives a ratio $\Delta/k_bT_c\approx1.5$, comparable to the BCS prediction. A slight underestimation of the Dynes fit to the measured spectra at higher bias voltages may suggest a small anisotropy of the superconducting gap. Crucially, however, there is no evidence of nodes in the superconducting order parameter or of in-gap states. Similar conclusions can be drawn from measurements performed using a superconducting tip (Fig.~\ref{f:STM}(c)), where again a clear lack of any zero-energy bound states is evident. 

Measuring the collapse of the superconducting gap in a  magnetic field applied normal to the sample surface (see Supplemental Fig.~5~\cite{sup} for a detailed description), we estimate an upper critical field, $H_{c2}^\perp \approx 20$~mT. This is consistent with the upper critical field of the bulk superconducting state as judged from susceptibility measurements~\cite{leng_type_2017}. Intriguingly, the bulk superconductivity in PdTe$_2$ has recently been reported to be of type I character~\cite{leng_type_2017}. We have, however, observed a vortex core in our STM measurements in a magnetic field of 7$\pm$2~mT (Fig.~\ref{f:STM}(d)), indicating that the superconductivity we probe is of type-II character. 

From the measured decay length of the vortex core of $175 \pm 67$~nm  (Fig.~\ref{f:STM}(d)), this would suggest an upper critical field of $H_{c2}\approx 11$~mT for a BCS superconductor, assuming a one-to-one correspondence of the measured decay length to the coherence length. This is in reasonable agreement with our measured value of the surface upper critical field. Finally, we note that scanning tunneling spectroscopy performed as a function of distance away from the centre of the vortex core (Fig.~\ref{f:STM}(e)) again yields no evidence for the presence of zero-energy bound states. Taken together, these results demonstrate that the surface of PdTe$_2$ supports a conventional fully-gapped $s$-wave superconducting state, well described by BCS theory. This co-exists with well-defined topologically non-trivial surface states, with complex multi-component and multi-valley Fermi surfaces and rich vorticial spin textures. 

The complexity of the surface Fermi surface, as well as the presence of numerous bulk states degenerate in energy but at different in-plane momenta, may ultimately explain why the dominant superconducting pairing remains topologically trivial at the surface. In any case, our results clearly demonstrate that the presence of topologically non-trivial states at the Fermi level is not a sufficient criterion to realise topological superconductivity. Beyond this, our findings highlight the importance of $k_z$-dependent band inversions within a single orbital manifold for generating topological surface states with rich and complex surface Fermi surfaces. Moreover, they  demonstrate how these can be effectively tuned by varying interlayer hopping strengths, paving the way to the design of new topological materials.

\

\begin{acknowledgements}
\noindent {\it Acknowledgements:} 
We gratefully acknowledge support from the Leverhulme Trust, the Engineering and Physical Sciences Research Council, UK (Grant Nos.~EP/M023427/1 and EP/I031014/1), the Royal Society, CREST, JST (Nos. JPMJCR16F1 and JPMJCR16F2), the Japan Society for Promotion of Science (Grant-in-Aid for Scientific Research (S); No. 24224009 and (B); No. 16H03847), the International Max-Planck Partnership for Measurement and Observation at the Quantum Limit and TRF and SUT Grant No. BRG5880010. This work has been partly
performed in the framework of the nanoscience foundry and fine analysis (NFFA-MIUR
Italy, Progetti Internazionali) facility. OJC, MJN, LB, VS, and JMR acknowledge EPSRC for PhD studentship support through grant Nos.~EP/K503162/1, EP/G03673X/1, EP/L505079/1, and EP/L015110/1. IM acknowledges PhD studentship support from the IMPRS for the Chemistry and Physics of Quantum Materials. We thank Diamond Light Source (via Proposal Nos.~SI9500, SI12469, SI13438, and SI16262) and Elettra synchrotron for access to the I05 and APE beamlines, respectively, that contributed to the results presented here. The research data supporting this publication can be accessed at Ref.~\cite{metadata}.
\end{acknowledgements}

\end{document}